\documentclass[traditabstract]{aa} 
\usepackage{graphicx}
\usepackage{txfonts}
\usepackage{natbib}
\bibpunct{(}{)}{;}{a}{}{,} 
\begin{document}
   \title{A high-resolution three-dimensional model of the solar
  photosphere derived from Hinode observations}

   \author{H. Socas-Navarro\inst{1, 2}
   }

   \institute{Instituto de Astrof\'\i sica de Canarias,
     Avda V\'\i a L\'actea S/N, La Laguna 38205, Tenerife, Spain
   \and
   Departamento de Astrof\'\i sica, Universidad de La Laguna, 38205, 
   La Laguna, Tenerife, Spain 
 }

   \date{Received September 15, 1996; accepted March 16, 1997}

   \authorrunning{Socas-Navarro}
   \titlerunning{An empirical 3D model of the solar photosphere}

\newcommand {\FeI} {\ion{Fe}{i}}
\newcommand {\ltau} {$\log(\tau_{5000})$}

\abstract{
  A new three-dimensional model of the solar photosphere is presented
  in this paper and made publicly available to the community. This
  model has the peculiarity that it has been obtained by inverting
  spectro-polarimetric observations, rather than from numerical
  radiation hydrodynamical simulations. The data used here are from the
  spectro-polarimeter onboard the Hinode satellite, which routinely
  delivers Stokes $I$, $Q$, $U$ and $V$ profiles in the 6302~\AA \
  spectral region with excellent quality, stability and spatial
  resolution (approximately 0.3''). With such spatial resolution, the
  major granular components are well resolved, which implies that the
  derived model needs no micro- or macro-turbulence to properly fit
  the widths of the observed spectral lines. Not only this model fits
  the observed data used for its construction, but it can also fit
  previous solar atlas observations satisfactorily.
}

   \keywords{ Sun: abundances -- Sun: atmosphere -- Sun: granulation
     -- Sun: photosphere -- Stars: abundances -- Stars: atmospheres 
}

   \maketitle
%

\section{Introduction}

The controversial debate sparked in recent years regarding the
chemical abundances of the Sun and other stars, and the profound
consequences that the proposed revision would have in many areas of
astrophysics, has bluntly put forward an alarming weakness in the very
foundations of our field. In particular, this issue has revealed
the pivotal importance of the choice of a suitable model atmosphere in
the process of abundance determination. Traditional one-dimensional
(1D) models of the solar atmosphere were produced empirically by
adjusting their parameters to reproduce more or less adequately a number
of observables, including spectral lines and continua. Some of the
most widely used empirical 1D models are the Harvard-Smithsonian
Reference Atmosphere (HSRA, see \citealt{GNK+71}), or those of
\cite{HM74}, \cite{VAL81}, \cite{FAL93}, etc.

A new generation of three-dimensional (3D) models, developed
theoretically {\em ab initio} from radiation hydrodynamical
simulations was used in a series of papers to propose a revision of
the solar chemical composition leading to a significantly lower
metalicity (see, e.g., \citealt{GAS07} and references therein). Of
particular importance has been the issue of the oxygen abundance
raised by \cite{AGS+04}, prompting what some authors have dubbed the
{\em solar oxygen crisis} \citep{APK06} and followed by controversy on
whether the proposed revision should be adopted or not
(e.g., \citealt{BBP+05,SNN07,CSN08,BA08,A08,SAG+09}). To add even more
confusion, another theoretical 3D model was used by \cite{CLS+08} to
derive the solar oxygen abundance, resulting in higher values than
those of \cite{AGS+04}. 

\cite{APK06} points out that the relative merits by which one measures
the success of a 3D theoretical model and those of an empirical 1D
one are certainly different. They suggest that while 3D is obviously
to be preferred over 1D, an empirical model is more suited than a
theoretical one for abundance determinations and it is not clear how
these two factors balance out in the trade off. 

The path taken in this work is an attempt to combine the best of both
approaches by deriving a 3D model from observations. This is possible
now for two reasons mainly:1)we have instrumentation with the
capability of providing detailed spectra with sufficient spatial
resolution to resove the granular motions in the solar photosphere;
and 2)because the analysis techniques (e.g., inversions) are mature
enough that it is computationally affordable to undertake a project
that involves the detailed study of a very large set of profiles.

Nonwhitstanding the focus of the present discussion on chemical
abundance determinations, an empirical 3D model is also of potential
usefulness in a wide range of investigations. Noteworthy examples are
aiding in the development and verification of theoretical {\em ab
  initio} models, predicting the expected shapes of continua and lines
and, since in this case it also includes the photospheric magnetic
field, their Zeeman polarization signatures as well.

The model presented here is publicly available and may be downloaded
both as an IDL savefile or in raw binary format\footnote{
As a courtesy, potential users are kindly requested to contact the
author explaining the nature of their investigations and the intended
use of the model.
} {from the following URL: 
\begin{verbatim}
ftp://download:data@ftp.iac.es/
\end{verbatim} The files are licensed under the GPLv3 general public
  license\footnote{See http://www.gnu.org/licenses/gpl.html}
  which explicitly grants permission to copy, modify (with
  proper credit to the original source and explanation of the modifications) and
  redistribute the software.
}

\section{Observations and data reduction}
\label{sec:observations}

The dataset used in this work was acquired starting at UT 19:32:10 on
2007 September 24 with the spectro-polarimeter (SP) of the solar
optical telescope (SOT) onboard the Hinode satellite
(\citealt{Hinode1,Hinode2,Hinode3,Hinode4,Hinode5}). The observed
field of view was very close to disk center, at heliocentric
coordinates (-16, -6) arc-seconds. The slit stepping projected on the
solar disk was of 0.15'' and its length of 162'' with a cadence of
13~s (exposure time was 12.8~s). A total of 1024 slit positions were
recorded, resulting in a total field of view of 151''$\times$162''
(although only a smaller sub-field will be analyzed in detail here, as
explained in Sect.~\ref{sec:analysis} below).

As usual, the spectral coverage of the Hinode SP has 112 wavelength
samples with a pixel sampling of 21.4~m\AA \, and spanning the range
between 6300.89 and 6303.27~\AA . The absolute wavelength calibration
was obtained by comparing the average spectrum to the Kitt Peak
Fourier Transform Spectrometer (FTS) disk center intensity atlas of
\cite{NL84}. With this exposure time, the signal-to-noise ratio in the
data, measured as the standard deviation of the Stokes Q, U and V
profiles in the continuum, is of 1.2$\times$10$^{-3}$ in units of the
continuum intensity. All of the Stokes parameters exhibit
approximately the same amount of noise.

\begin{figure}
  \centering
  \includegraphics[width=9cm]{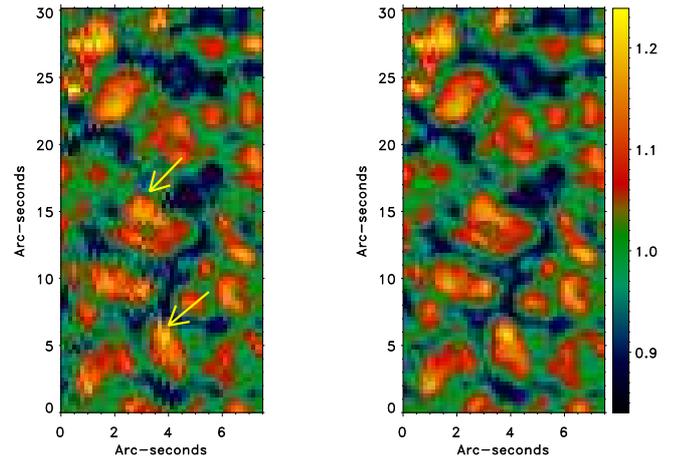}
  \caption{Detail of the {continuum intensity} in the observed
    field of view showing some glitches due to the sudden displacement
    of the frame along the slit direction {(the arrows point to
      some of the most obvious examples)}. Left panel: Original
    image. Right panel: Corrected as explained in the text.}
  \label{fig:reduc}
\end{figure}

All the data were first processed from Level~0 to Level~1 using the
standard Hinode SOT/SP data reduction pipeline. The resulting Level~1
data were then subject to a number of additional steps, as follows:
\begin{itemize}
\item Remove wavelength offset in the slit stepping
  direction: The absolute wavelength calibration was obtained for the
  average spectrum. However, if one computes the line minimum position
  for each $(x,y)$ point (with $x$ being the slit stepping direction
  and $y$ the direction along the slit), average along the
  $y$-direction and examine its variation with $x$, a fluctuating
  pattern appears. The $y$-averaged line center varies by as much as
  11 pixels (from minimum to maximum) in the $x$ direction. This
  variation is measured consistently in both \FeI \ lines at 6301.5
  and 6302.5 \AA \ (the same value is obtained for both lines, with
  the maximum difference over the whole range being 0.02 pixels). To
  remove this undesirable effect, the line center displacement
  averaged over the two lines and the $y$-direction is subtracted by
  reinterpolating the four Stokes profiles at each spatial
  position. As mentioned above this is done over the full original
  field of view of 1024$\times$1024 pixels to ensure that we have as
  much statistics as possible in the spatial average along the slit.
\item Remove pointing jumps in the slit direction: Visual inspection
  of the continuum maps shows some locations with obvious local
  discontinuities, caused by the slit moving a few pixels up or down
  in the $y$-direction. This glitch occurs relatively infrequently but
  has been corrected by comparing each pixel with its surroundings in
  the $x$-direction (3 pixels to the left and 3 pixels to the right)
  and testing if shifting the slit by up to 10 pixels up or down would
  improve continuity in the continuum map. The optimal continuity
  shift found in this manner is then averaged along the $y$-direction
  (again taking the full 1024 pixels to maximize statistics) and
  applied to the data. The results of this procedure may be seen in
  the example shown in Fig.~\ref{fig:reduc}.
\item Following \cite{DGL+08}, a parasitic stray light contamination of
  5\% is removed from all the intensity spectra before normalizing all the
  dataset to the average continuum intensity over the entire field of view.
\end{itemize}

After all data processing, a few flatfield artifacts still remain
visible in the continuum images. However, the smaller subfield
selected for further analysis below has been selected to avoid the
affected areas. The resulting rms granulation contrast in the region
is 7.7\% of the mean continuum value.

\section{Analysis}
\label{sec:analysis}

\begin{figure*}
  \centering
  \includegraphics{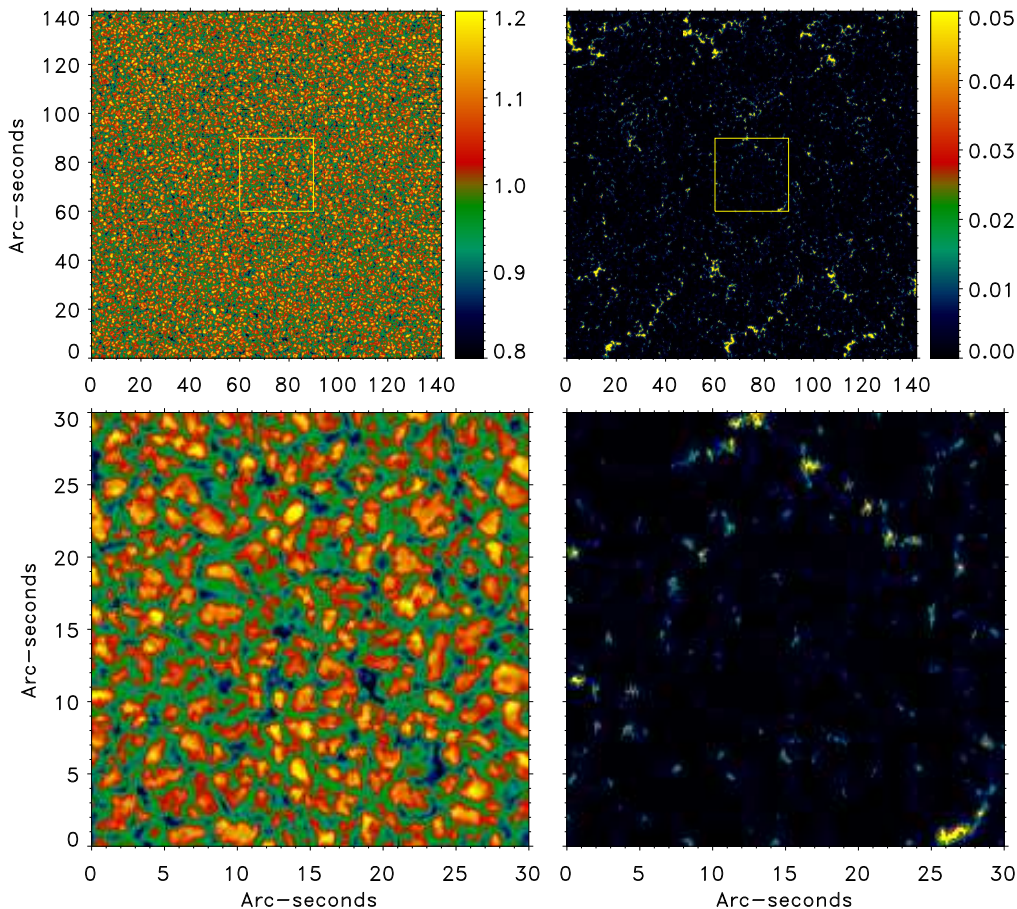}
  \caption{Full field of view (upper panels) and subfield selected for
    inversions (lower panels). Left: Continuum image. Right: Synthetic
    magnetogram, saturated at 5\% of the average continuum
    intensity. The selected subfield avoids some flatfield artifacts
    that are visible as horizontal segments in the upper left
    panel.
}
  \label{fig:fields}
\end{figure*}

For practical reasons, a smaller subfield of view of 200$\times$200
pixels has been selected for further analysis. This size is large
enough to include abundant statistics on granulation and quiet Sun
fields (including both network and internetwork) and at the same time
small enough to allow a full inversion of each spatial point
individually in a reasonable time span using a
supercomputer. Figure~\ref{fig:fields} shows continuum intensity and a
synthetic magnetogram (constructed simply as the absolute value of the
circular polarization signal integrated over a 100~m\AA \, range in the
blue lobe of both lines) of the original full field of view and the
subfield selected for inversion.

The inversions were carried out with the code NICOLE
\citep{SNdlCAR+10}, which is an improved implementation of the
original NLTE inversion code of \cite{SNTBRC00a}. It makes use of
response functions similarly to the popular LTE code SIR of
\cite{RCdTI92}, but has some features that are important for our
purposes here. NICOLE is supported by a very wide variety of platforms
and has built-in MPI parallelization, which makes the use of
supercomputers rather simple and straightforward. Moreover, it 
has NLTE capabilities, including an option that is used for part of
this work (see Sect.~\ref{sec:depcoef}) in which it is possible to
supply a (fixed) set of departure coefficients for the upper and lower
levels of the spectral lines.

The calculations presented in this paper have been performed using the
LaPalma supercomputer of the Instituto de Astrof\'\i sica de
Canarias {which harbors 512 64-bit processors, running at a speed of
2.2~GHz. A typical run, including the various passes described below,
requires about 6 hours on 50 CPUs. }


The observed spectral region contains two prominent \FeI \ lines (at
6301.5 and 6302.5~\AA ) whose profile shapes are fitted by the
inversion code to find the optimal depth stratification of the
atmospheric variables. These lines are well described with a LTE
treatment (the effects of NLTE corrections will be analyzed in
Sect.~\ref{sec:depcoef} below). The atomic parameters are taken from
the VAL-D database \citep{PKR+95}, except for the transition
probabilities. For the 6301.5~\AA \, line we use the laboratory
measurements of \cite{BKK91}.

Unfortunately, no similar measurements exist for the 6302.5~\AA \,
line and we derived it empirically in the following manner. We first
inverted the 6301.5~\AA \, line in the FTS atlas mentioned above. With
the model thus obtained we synthesized the 6302.5~\AA \, line
adjusting the $\log(gf)$ parameter until a satisfactory fit to the
observations was attained. Collisional broadening is treated with the
formalism of \cite{AOM95} with the parameters $\alpha$ and $\sigma$
obtained using the code of \cite{BAOM98}. The atomic parameters used
are given in Table~\ref{table:linedata} ($r_0$ denotes de B\"ohr
radius).

\begin{table}
  \caption[]{Spectral line data}
  \label{table:linedata}
  $$ 
  \begin{array}{ccccccc}
    \hline
    \noalign{\smallskip}
  \lambda \  (\mbox{\AA})  & Excitation  & log(gf) & Term & Term & \sigma 
  &  \alpha \\
                   &  pot. \ (eV) &         & (lower) & (upper)& (r_0^2) & \\
    \noalign{\smallskip}
    \hline
    \noalign{\smallskip}
    6301.5 & 3.654 & -0.718  & ^5P_2 & ^5D_2 & 834.4 & 0.243   \\
    6302.5 & 3.686 & -1.160  & ^5P_1 & ^5D_0 & 850.2 & 0.239   \\
    \noalign{\smallskip}
    \hline
  \end{array}
  $$ 
\end{table}

For each pixel in the 200$\times$200 subfield, the inversion procedure
takes a starting guess model atmosphere and iteratively modifies it by
adding a correction (which, in general, is depth-dependent) to it,
seeking the best fit to the observations with the synthetic profiles
computed from that modified model. The correction that is applied to
the guess model at each iteration may be constant with height
(depth-independent), linear with $\log(\tau_{5000})$ (the logarithm of
the continuum optical depth at 5000 \AA ) or it can be constructed as
a concatenation of linear segments spanning the whole depth range. In
this last case, the number of segments to use may be set arbitrarily
by the user. In this manner, depending on the amount of information
available, we can decide how much detail of the depth stratification
we wish to retrieve. One needs to reach a compromise to adjust the amount
of freedom in the model to the information available in the data. Too much
freedom results in degeneracies and leads to uniqueness issues and
other complications. On the other hand, being too restrictive results
in poor fits and not extracting all of the available
information. Usually, some experimentation with a few test cases is
very useful to determine a nearly optimal set of parameters.

The choice of freedom in each physical parameter of the model
(temperature, magnetic field, velocities, etc) is made in practice by
selecting the number of inversion nodes \citep{RCdTI92}, which are
actually the free parameters in the code. Using one node we have a
depth-independent correction for the entire atmosphere. With two nodes
we produce a correction that scales linearly with
$\log(\tau_{5000})$. Similarly, with three or more nodes we produce
more complicated variations. Following the recommendation of
\cite{RCdTI92}, we proceed in two successive cycles to improve
convergence. We start in the first cycle with a relatively small
number of nodes to obtain a first approximation to the solution,
fitting the overall shape of the profiles. We then increase the number
of nodes to obtain a better solution allowing more freedom to fit more
subtle properties, such as line asymmetries.

In addition to these cycles, NICOLE implements a
multiple-initialization scheme in which the inversion (both cyles) is
repeated a number of times randomizing the initial guess. The process
is stopped when a good fit is achieved or when a preset maximum number
of inversions have been done. The criterion to decide what we consider
a good fit and the maximum number of inversions are both defined by
the user. This is helpful when, as in this case, one is batch
processing a large number of profiles to minimize the risk of the
algorithm settling in a secondary minimum. The price to pay for the
ensuing improvement in stability is, of course, more CPU time. In the
calculations presented here, the stopping criteria are: a)a fit better
than 1\% of the profile on average or b)up to five inversion attempts.

As a starting guess we take the HSRA model. Two different approaches
are taken depending on whether the profile to invert exhibits
significant polarization or not. In the first case, which accounts for
34\% of the profiles, we consider two atmospheres coexisting
side-by-side in the pixel, one magnetic and the other non-magnetic. To
determine the non-magnetic component, we start with a preliminary
inversion of Stokes~$I$ only (weights for $Q$, $U$ and $V$ are set to
zero) without any magnetic field. In this inversion we allow for some
micro-turbulent line broadening to improve the fit since it is
unlikely that the external atmosphere will be entirely
field-free. However, we do not invert for a magnetic field here since
our main focus is the field in the second component which is the one
that gives rise to the Stokes $Q$, $U$ and $V$ profiles. The resulting
model and Stokes~$I$ profile from this preliminary inversion are then
taken as the (fixed) external non-magnetic atmosphere in a subsequent
inversion in which the filling factor is a free parameter. In the case
that the pixel considered does not exhibit Stokes signal we proceed
with a simple one-component inversion. The number of nodes employed in
both cases is shown in Table~\ref{table:nodes}. As mentioned above,
only the first (preliminary) inversion has microturbulence so that the
code is able to handle the extra broadening due to the magnetic
field. In the second and third inversions, where the magnetic field is
taken into account, microturbulence is set to zero.

\begin{table*}
  \caption[]{Inversion nodes}
  \label{table:nodes}
  $$ 
  \begin{array}{lccccccccc}
    \hline
    \noalign{\smallskip}
     & \vline &         & Magnetic \, pixel &        &    &     & \vline
     &  Non-magnetic \, pixel &  \\
   Physical  & \vline &  Inversion 1 &   & \vline &Inversion 2  &    & \vline &        \\
   Parameter  & \vline & Cycle 1 & Cycle 2 & \vline& Cycle 1 & Cycle 2 & \vline & Cycle 1 & Cycle 2 \\
    \noalign{\smallskip}
    \hline
    \noalign{\smallskip}
  Temperature &        &  3  &  5  & &     2   &   3  & &  3 &  5 \\  
  L.o.s. \,  velocity    &        &  1  &  3  & &     1   &   2  & &  2 &  3 \\     
  Microturbulence &    &  1  &  1  & &     0   &   0  & &  0 &  0 \\
  B_{long}     &        &  0  &  0  & &     1   &   2  & &  0 &  0 \\
  B_x         &        &  0  &  0  & &     1   &   2  & &  0 &  0 \\
  B_y         &        &  0  &  0  & &     1   &   2  & &  0 &  0 \\
  Filling \, factor &     &  0  &  0  & &     1   &   1  & &  0 &  0 \\
    \noalign{\smallskip}
    \hline
  \end{array}
  $$ 
\end{table*}

In the table, $B_{long}$ refers to the longitudinal (i.e., along the
line-of-sight) component of the magnetic field, whereas $B_x$ and
$B_y$ are the transverse components, projected on the plane of the
sky. The $x$-direction is the field azimuth reference, defined by the
plane of vibration of light with $Q$$>$0 and $U$=0 in the calibration
pipeline.  The synthetic Stokes spectra computed at each step of the
iteration are convolved with the instrumental profile of the Hinode SP
(Lites, private communication) to make sure that synthetic and
observed profiles are comparable.

A final pass is performed restarting the inversion with a smoothed
version of the result to fix a small percentage of pixels that did not
converge properly. The smoothing is done with a 3$\times$3 median box,
but excluding the central point (i.e., the pixel to be re-inverted).

\begin{figure*}
  \centering
  \includegraphics[width=19cm]{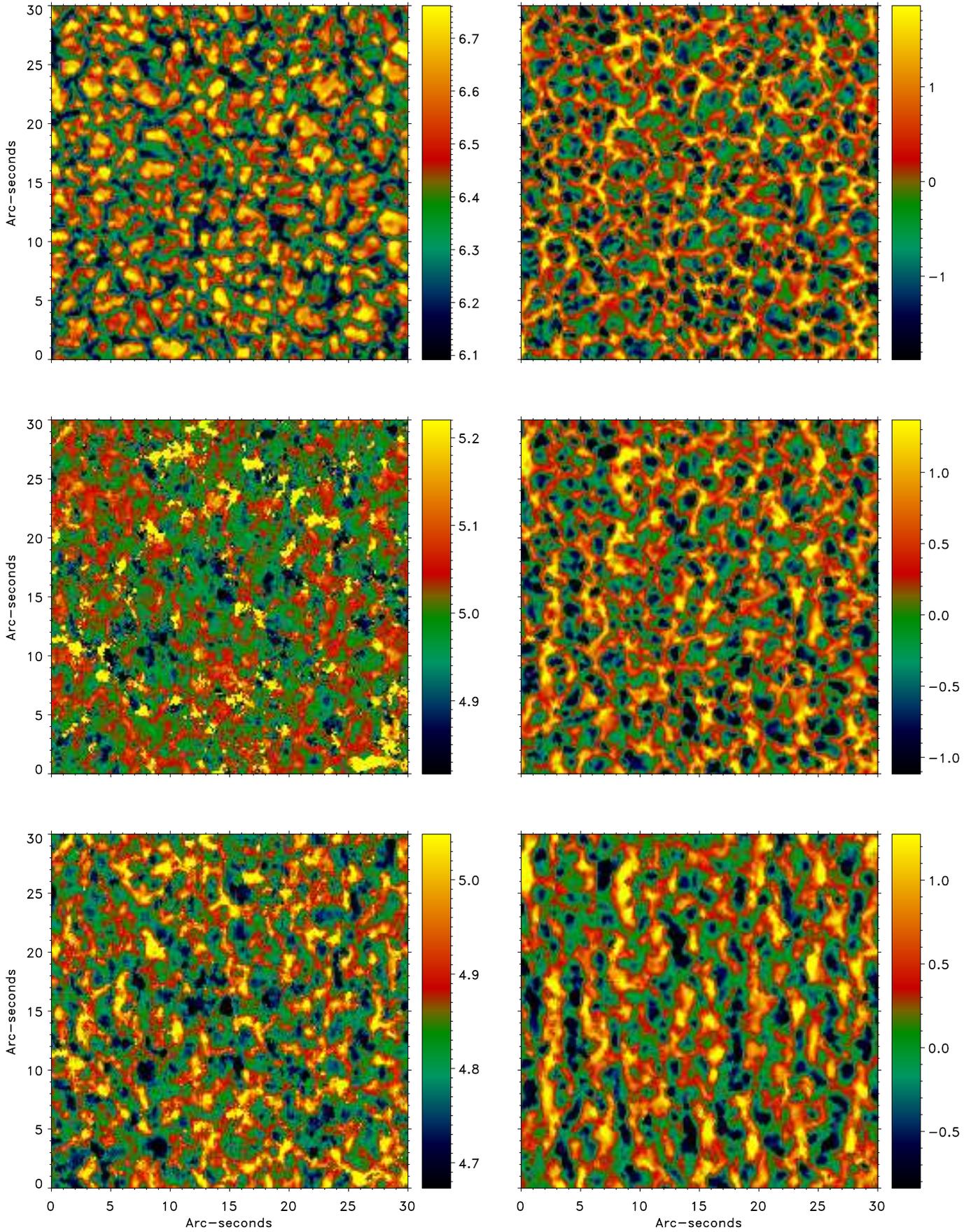}
  \caption{Horizontal cuts of the temperature (left) and line-of-sight
    velocity (right) at $\log(\tau_{5000})=0$ (top), $-1$ (middle) and
    $-2$ (bottom). Units are kK for temperature and km~s$^{-1}$ for
    velocity. The astrophysical convention, where positive velocities
    indicate redshift (downflows), is adopted throughout this paper.
  }
  \label{fig:maps1}
\end{figure*}

A sample of the resulting cubes can be seen in Fig.~\ref{fig:maps1}
which shows horizontal (in $\tau_{5000}$) cuts of the temperature and
line-of-sight velocity at three representative heights in the
photosphere. For the two-component inversions (pixels with magnetic
signal), an average of the internal and external atmospheres weighted
with their respective filling factors is shown.

The figure shows the characteristic convective motions at the base of
the photosphere (top panels), with hot granules upflowing and cool
lanes downflowing. The flow structure does not change too much until
we reach the upper photosphere (bottom panel) in which the granulation
pattern starts to dissolve leaving structures that vaguely appear to
be vertically oriented. Since there is no preferred direction in the
field of view, such structures are likely the result of observing a very
dynamic upper atmosphere through a vertical slit. The temperature
stratification starts with the granulation at the bottom but at
\ltau$=-1$ is dominated by hot patches in the magnetic areas embedded in
a more uniform non-magnetic background. Finally, at the \ltau$=-2$
height we start to see reversed granulation, where the
photospheric granules are now cooler and the lanes are hotter.

\subsection{NLTE corrections}
\label{sec:depcoef}

The \FeI \ lines used in this work are affected by (generally very
small) NLTE effects \citep{STB01}. Both the upper and lower levels of
the transitions are slightly underpopulated with respect to their LTE
values. This has virtually no effect on the source function, which
goes with the ratio of the level populations, but produces a small
overall decrease in the line opacity. 

We introduce a correction for NLTE effects in the inversion scheme by
using some departure coefficients computed by \cite{STB01} in the 3D
hydrodynamic model of \citet[][see also \citealt{SN98}]{ALN+00}. We
take the run with \ltau \ of the departure coefficients for both
levels of the 6301.5 and 6302.5 \AA \ transitions computed by
\cite{STB01} in a typical granule and a typical lane in the 3D
simulation. These values are then assigned to a particular granule and
a lane pixel in the Hinode observations. For all other pixels we carry
out a linear interpolation based on the continuum intensity. The
departure coefficients obtained are kept fixed during the inversion,
even though the atmospheric parameters change. While this method is
not exact, it is a good approximation to correct for an effect that is
already small anyway. A full NLTE solution for the Fe atom would be a
tremendous undertaking and a full 3D inversion such as the one
presented here would not be possible in practice due to the processing
power required. With our approximation, the departure coefficients may
be inconsistent with the starting guess atmosphere but as the
inversion proceeds and approaches the solution, they become gradually
more realistic.

\section{Results}

\begin{figure*}
  \centering
  \includegraphics[width=19cm]{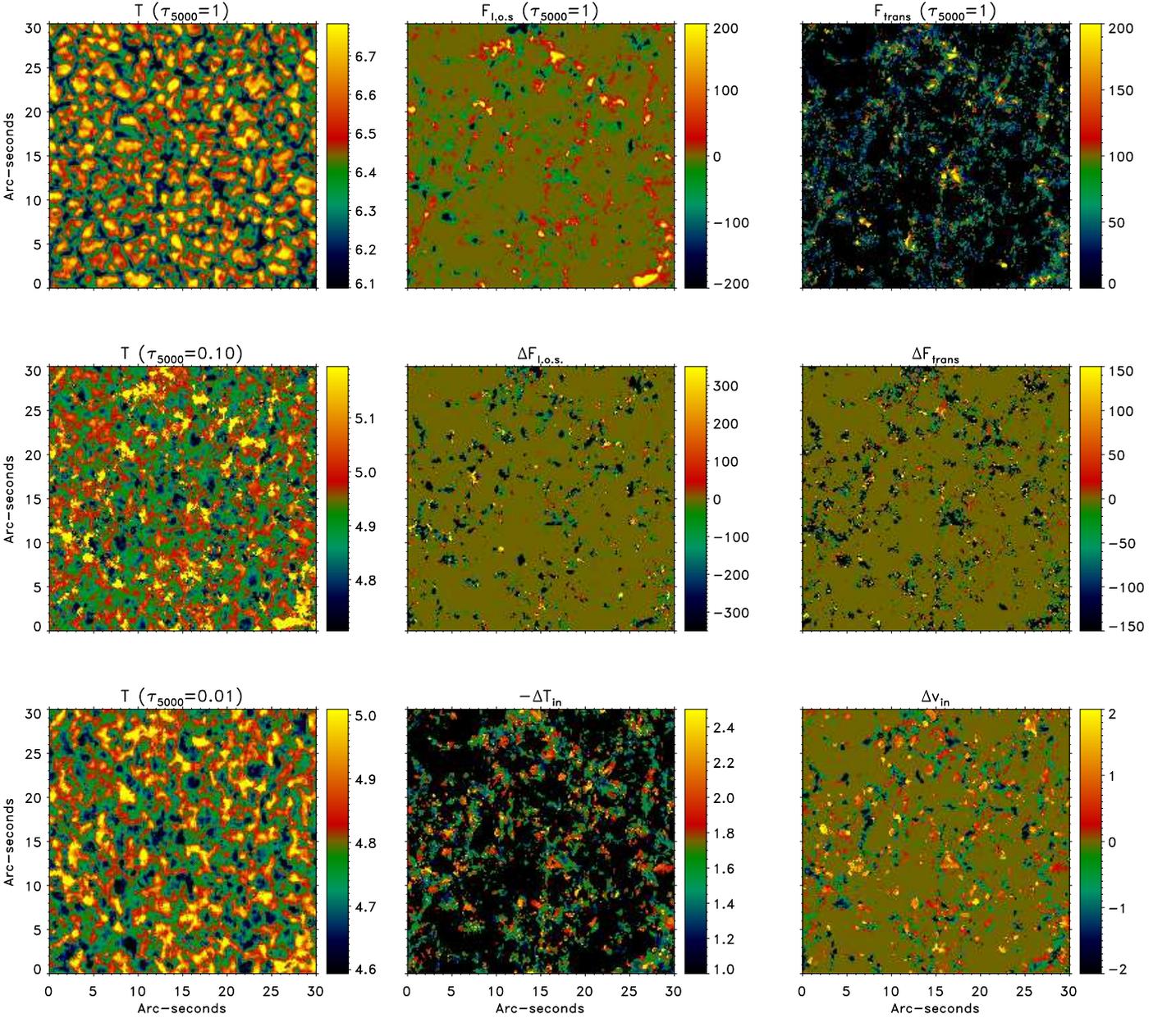}
  \caption{Horizontal (in the $\tau_{5000}$ depth scale) cuts of
    several interesting physical parameters in the multi-cube
    resulting from the inversion, including NLTE correction. The left
    column shows the temperature in kK at three different heights. The
    top row also shows the line-of-sight magnetic flux density (middle
    panel) in G and the transverse (on the plane of the sky) magnetic
    flux density (right panel) at the base of the photosphere. The
    middle row shows the variation of the flux density from \ltau$=0$ to
    \ltau$=-1$ (middle panel: line-of-sight component; right panel:
    transverse component). The bottom row shows the variation of the
    temperature in kK (middle panel) and the line-of-sight velocity in
    km~s$^{-1}$ (right panel) from \ltau$=0$ to \ltau$=-1$ inside the
    magnetic atmosphere.}
  \label{fig:maps2}
\end{figure*}

With the NLTE correction, the cores of the lines are fitted better and
therefore the results for the upper layers are less noisy (see
Fig.~\ref{fig:maps2}; compare bottom-left panel to that in
Fig.~\ref{fig:maps1}). Figure~\ref{fig:maps2} shows also the magnetic
flux density in the region, revealing many internetwork field
structures. The transverse flux density is more noisy because the
Stokes~$Q$ and~$U$ signals are weaker than Stokes~$V$ and
correspondingly more difficult to detect. Nevertheless, we can see
that a large fraction of the area harbors fields with a measureable
horizontal component, in agreement with recent results such as those
of \cite{LKSN+08}.

The magnetic field height gradient is mostly negative as one would
expect. However, some concentratios show an increase of the field
strength with height. Temperatures inside the magnetic atmospheres
always decrease with height but there is a wide range of variation of
$\sim$1~kK among various pixels. The mass flows may be upward or
downward directed. However, the stronger flux elements are associated
with downflows. If we consider pixels with a longitudinal magnetic
flux stronger than 100~G, 84\% harbor downflows and only 16\% are
upflowing. The external (non-magnetic) atmosphere is also undergoing
downflows in those pixels.

\begin{figure}
  \centering
  \includegraphics[width=9cm]{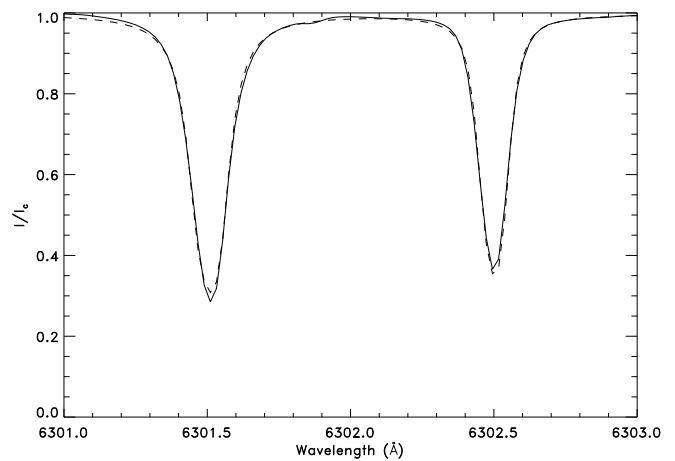}
  \caption{Average intensity spectrum from the Hinode SP observed dataset
    (solid) and from the 3D model (dashed). Spectra are normalized to
    the continuum intensity ($I_c$).
}
  \label{fig:hinodefits}
\end{figure}

Generally speaking, the fits to the individual profiles are
good. Figure~\ref{fig:hinodefits} shows the average intensity profile
in the observed region and the average synthetic profile from the
model. Notice that this is not actually a fit but an average of the
200$\times$200 individual fits.

\begin{figure}
  \centering
  \includegraphics[width=9cm]{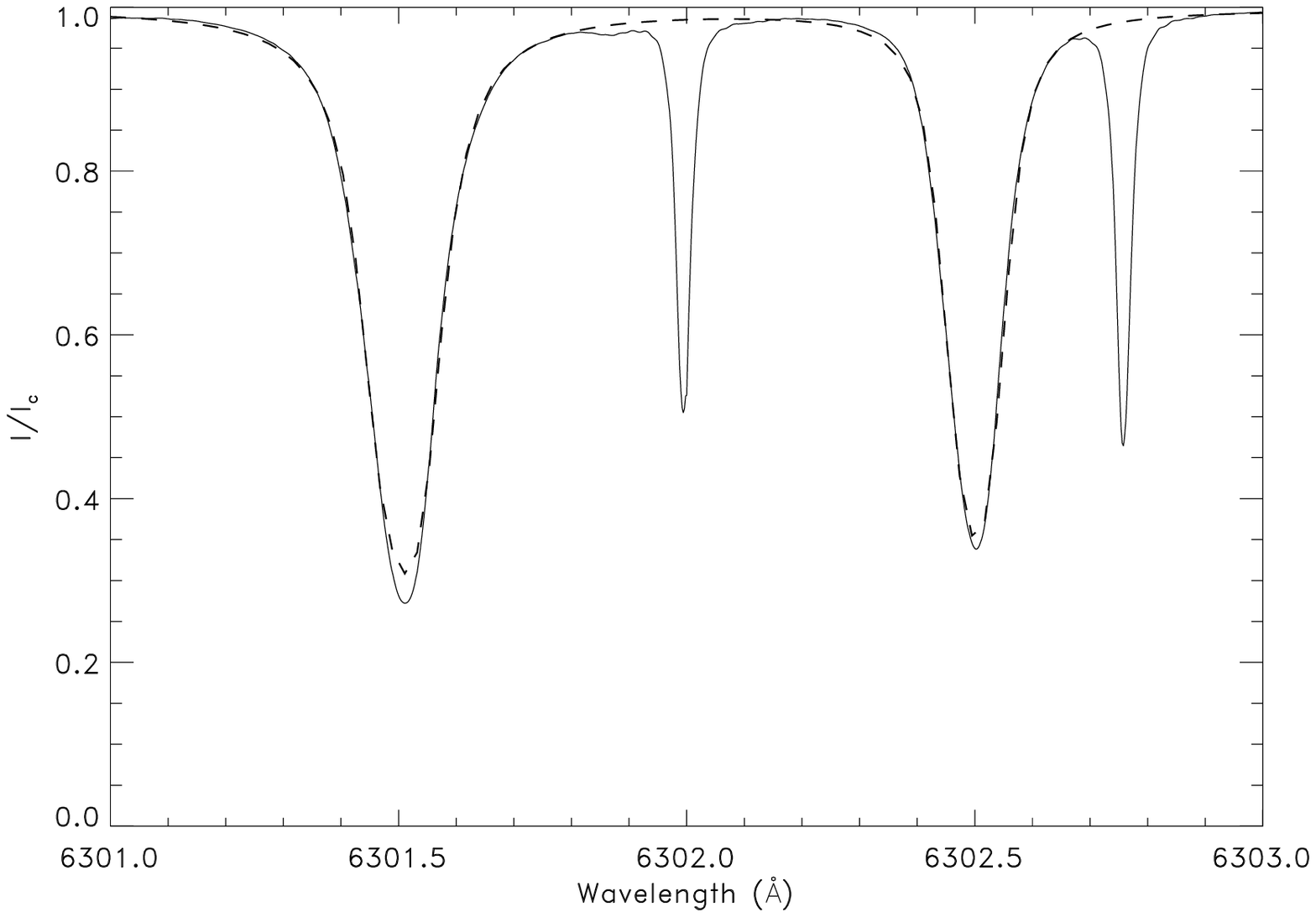}
  \caption{FTS atlas intensity spectrum (solid) and average profile
    synthesized from the 3D model (dashed) without instrumental
    profile (i.e., with very high spectral resolution). The lines at
    6302.00 and 6302.76 \AA \ are telluric lines (absent in the Hinode
    data). Spectra are normalized to the continuum intensity ($I_c$).
}
  \label{fig:ftsfits}
\end{figure}

We can also test the model against atlas observations. If we
synthesize the profiles in the absence of macroturbulence and
instrumental profile (i.e., with very high spectral resolution), we
should obtain a profile very similar to that of an intensity atlas. In
Fig.~\ref{fig:ftsfits} we can see the comparison with the FTS atlas
\citep{NL84}. Overall, the model reproduces the atlas observations
with very high fidelity, including the line broadening. The absence of
microturbulence does not represent any problem for obtaining realistic
line shapes. This is not entirely surprising, since the observations
have sufficient spatial resolution to resolve the granular components
(granulation is the main contributor to the line broadening
traditionally characterized by microturbulence in 1D models). It is,
however, a noteworthy feature since the success in reproducing line
widths without requiring any microturbulence has been remarked as a
strong argument supporting theoretical 3D models derived in recent
years.

\begin{figure}
  \centering
  \includegraphics[width=9cm]{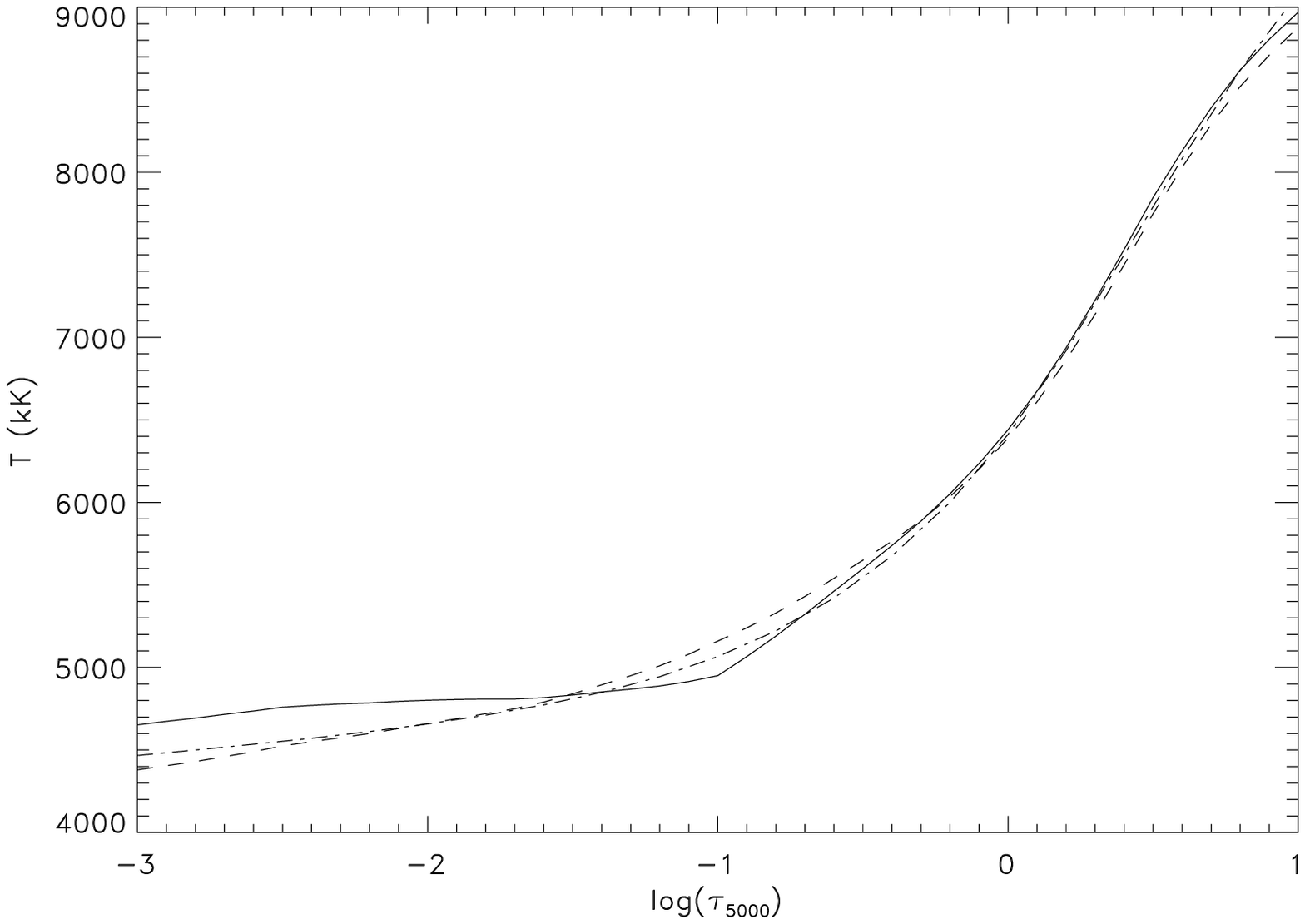}
  \caption{Comparison of the average temperature stratification in our
    3D model {(solid line)} to HSRA (dashed) and the average stratification of the
    \cite{AGS+04} 3D model {(dash-dotted)}.
}
  \label{fig:modelcomp}
\end{figure}

When we compare the average temperature stratification of our model to
others previously existing in the literature, we find that it is
similar, although slightly cooler in the middle photosphere (some
100{-200}~K at \ltau$=-1$) and warmer at the top ($\sim$150~K at
\ltau$=-3$). Figure~\ref{fig:modelcomp} compares the horizontally
averaged temperature to that of the Harvard-Smithsonian Reference
Atmosphere (HSRA, \citealt{GNK+71}) and to the average temperature of
\cite{AGS+04}. The HSRA is a semiempirical 1D model and one expects
some differences with the one presented here because in principle the
1D model that fits a given spectral dataset is not necessarily the
same as the average stratification of a 3D model that fits the same
dataset. In other words, the HSRA already incorporates an empirical
correction for 3D effects. The model of \cite{AGS+04} is a 3D model
but obtained from theoretical simulations. Compared to HSRA, it also
produces an elbow around \ltau$=-1$ but not as pronounced as our
model. In the upper layers, on the other hand, both HSRA and the
Asplund et al models are very similar in spite of the 1D vs 3D
difference in approach and both are cooler than the present model.

\begin{figure}
  \centering
  \includegraphics[width=9cm]{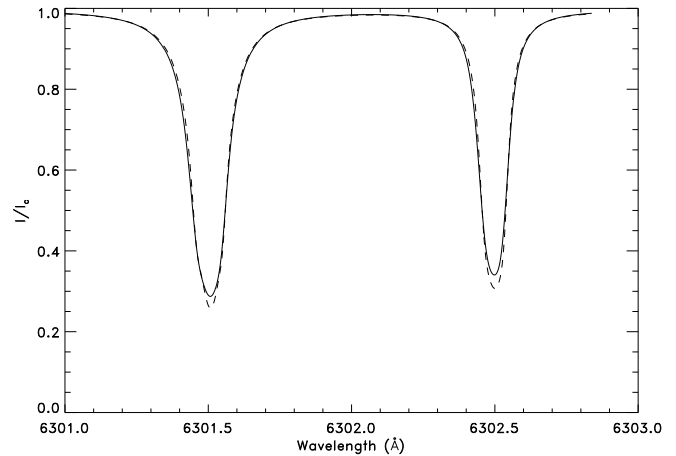}
  \caption{Comparison of the average spectrum from the 3D model
    (solid) to the spectrum produced by the average model
    (dashed).
}
  \label{fig:1Dvs3D}
\end{figure}

The importance of 3D effects has been emphasized in recent work (e.g.,
\citealt{AGS+04}) in the context of chemical abundance
determinations. To assess the importance of 3D effects in our model we
have compared the average emerging spectrum from what would be
obtained if one first averages the atmospheric stratification and then
compute the spectrum of such 1D average model. The results, displayed
in Figure~\ref{fig:1Dvs3D}, are in this case rather small (but
noticeable), with the main departure been at the core of the lines
(i.e., affecting the upper atmospheric layers). The lines computed in
1D are also slightly narrower, which goes in the direction of
compensating the change in equivalent width. In fact, the difference in
equivalent width between the 1D and 3D profiles is at the 1.5\% level.

\section{Uncertainties}

It is important to estimate the errors and uncertainties in the
derived model. Systematic errors may come from the normalization of
the Hinode data since the observations are not photometrically
calibrated. Instead, relative intensities to the disk-center average
continuum have been employed, with the subsequent assumption that such
disk-center average intensity is well reproduced by the HSRA
model. The visible continuum is very sensitive to temperature
variations, which has the advantage that it is relatively easy to
determine the photospheric temperature. On the downside, a small error
in the intensity calibration might have relatively large effects on
the temperature determination. For instance, a 1\% error in the
continuum calibration results in a $\Delta$T of between 150 and 200~K
(it is slightly different for granules and lanes). This example is
particulary pessimistic, as we do not expect such a large error in the
average intensity determination. Furthermore, the average temperature
of our model at $\tau_{5000}=1$ (6441~K) is consistent with values
from several other models existing in the literature, such as HSRA
(6390~K) or those of \citet[][6412~K]{AGS+04}, \citet[][6530~K]{HM74},
\citet[][6424~K]{VAL81} and \citet[][6520~K]{FAL93}.

\subsection{Sensitivity range}

\begin{figure}
  \centering
  \includegraphics[width=9cm]{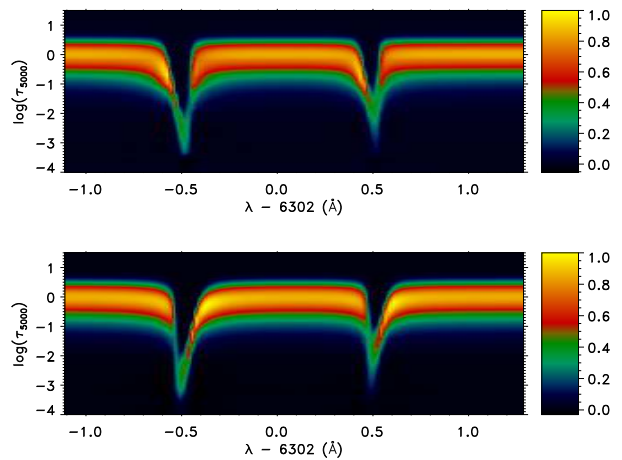}
  \caption{Normalized response functions $R(\lambda,\tau)$ to
    temperature in a granule (upper panel) and a lane (lower panel).
}
  \label{fig:rfs}
\end{figure}

Semi-empirical models derived from inversion of spectroscopic data,
such as this one, are reliable only in the atmospheric region where
the spectral lines exhibit some sensitivity to the physical
parameters. A suitable way to explore this range is by using the
so-called {\em response functions} \citep[e.g., ][]{RCdTI94}. Mathematically
speaking, the response function of a given Stokes parameter at a given
wavelength $I(\lambda)$ to perturbations in a given physical parameter
at a certain height in the atmosphere $T(\tau)$ is expressed as:
\begin{equation}
\label{eq:rf}
R(\lambda,\tau)= { dI(\lambda) \over dT(\tau) } \, .
\end{equation}
Here we use the same nomenclature to refer to the discrete form
(replacing the derivative with a ratio of finite increments) of
Eq~(\ref{eq:rf}) which, albeit not entirely equivalent from the
mathematical point of view, is very convenient in practical terms for
our study. 
The most straightforward way to compute a response function is by
brute force from its definition. We start with a reference atmosphere
from which the emerging spectrum is known and then perturb this
atmosphere, one point at a time, recompute the spectrum and calculate
the ratio $\Delta I(\lambda)/\Delta T(\tau)$ (assuming that
the perturbed variable is temperature at depth $\tau$). Response
functions to temperature in a granule and an intergranular lane are
displayed in Fig.~\ref{fig:rfs}.

From the response functions we can already see that the sensitivity
range of these two lines spans from roughly \ltau$=0$ to~$-3$,
approximately. Let us go a step further and use this information to
estimate the actual errors in the inversions. To this effect it is
important to consider not only how sensitive the lines are but also
how good a fit we have obtained. If we had only one wavelength, it is
straightforward to see that the error in the determination of a given
parameter (e.g., $T$ to fix ideas) may be approximated by  $\Delta T =
\Delta I / R$, 
where $\Delta I$ is the difference between the observed and the
synthetic observable and $R$ is the response function.

\begin{figure}
  \centering
  \includegraphics[width=9cm]{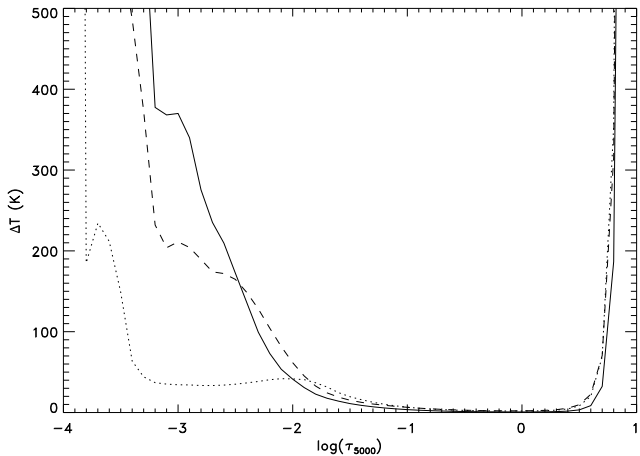}
  \caption{Estimated uncertainty in the temperature stratification for
    three different model atmospheres. Solid line: Brightest continuum
    location in the field of view (a granule). Dashed line: Darkest
    continuum location in the field of view (a lane). Dotted line: A
    location having a brightness equal to the average of the whole
    region. Coordinates on the lower panel maps of
    Fig.~\ref{fig:fields} are (14.5,8.2) for the granule, (12.0,25.5)
    for the lane and (22.6,8.5) for the average brightness point.
}
  \label{fig:errors}
\end{figure}

Consider now a case with many wavelengths and assume for simplicity
that the inversion result may be regarded as a weighted average of
individual measurements of $T(\tau)$, one at each wavelength, all
weighted by their individual response $R(\lambda,\tau)$. We can then
use the expression for the variance of a weighted mean to obtain in
our case:
\begin{equation}
  \label{eq:errT2}
  {1 \over \Delta T(\tau)^2  } = 
  \sum_{\lambda} { R(\lambda,\tau) \over \Delta I(\lambda) }
\end{equation}
Figure~\ref{fig:errors} shows the uncertainties in the temperature
determination obtained using this expression in three different pixels
of the field of view, which have been selected according to their
continuum brightness to include a granule (bright point), an
intergranular lane (dark point) and an average location. The residual
discrepancy between the average observed and synthetic spectra is
introduced in the equation as the $\Delta I(\lambda)$ function. The
figure shows that extremely small errors ($\sim$10~K or smaller) are
obtained for a fairly wide range of heights, between \ltau$=0.5$ and
-1.7, approximately. Up until \ltau$=-2.5$ we can still have relatively
small errors ($\sim$150~K) but above that height, measurements should
be regarded as highly uncertain. Notice, however, that the dependence
of the errors with the model atmosphere chosen is significant for the
upper layers. {The granule model is the one that has the error
  increasing more rapidly with height between \ltau$=-2$ and
  $-3$. In the lane model the error increases more
  slowly. Finally, }in the average atmosphere one can reach \ltau$=-3.4$
with very small errors of $\sim$50~K.

\subsection{Abundances}

For the calculations in this work we have employed the solar chemical
composition published by \cite{GS98}. Since the spectral lines
analyzed are from Fe transitions, this is the abundance value that
will have a stronger impact on our results. The 7.50 value in
\cite{GS98} seems to be very well established and, when NLTE effects
are considered (\citealt{STB01}), in good agreement with the
meteoritic value. The rest of the elements are relevant only for the
calculation of the background opacities.

To assess the impact of possible uncertainties in the abundances
employed, the synthetic profile produced from the average model was
inverted with different sets of abundances. A first test had the Fe
abundance fixed and a constant scaling was applied to all other
elements. The scaling factor ranged from 0.75 to 1.25. For the second
test the Fe abundance was varied between 7.40 and 7.60, while the rest
of the chemical composition was kept constant. 

\begin{figure}
  \centering
  \includegraphics[width=9cm]{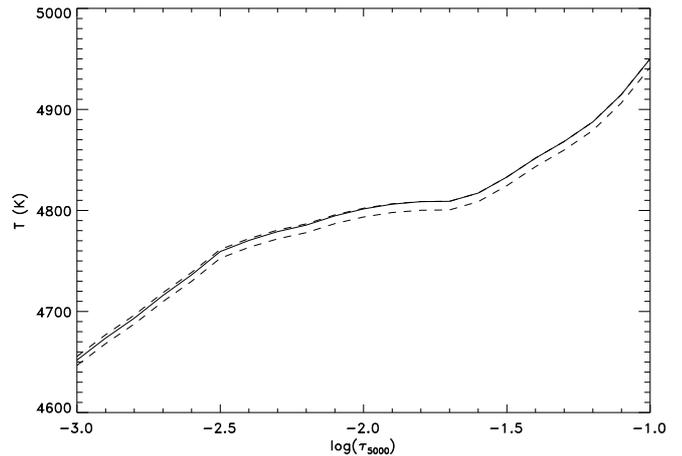}
  \caption{Models obtained when inverting the synthetic average
    spectrum with different values of the Fe abundance. The solid line
    is for the reference value of 7.50. The dashed lines are for 7.40
    (lower curve) and 7.60 (upper curve).
}
  \label{fig:abund}
\end{figure}

As expected, the only significant difference was found in the test
where the Fe abundance was modified. Figure~\ref{fig:abund} shows the
inversion with the reference value of 7.50 compared to the models
obtained with the interval extremes (7.40 and 7.60, respectively). The
changes are very small, mostly because the emerging intensity is
extremely sensitive to temperature in this range. This means that a
very small variation in the model temperature is able to compensate
for significant variations in the line parameters. The upside of this
conclusion is that semiempirical models such as the one presented here
are rather robust and not very sensitive to uncertainties in line
data, abundances, etc. Unfortunately, the downside is that abundance
determinations are extremely sensitive to the model employed.

\section{Conclusions}

The model presented {here} is novel in that it is
three-dimensional in nature and has been obtained from real
observations. It may hopefully serve as another platform for studies
of chemical composition (in which more 3D models are urgently needed),
to produce synthetic spectra (with or without polarization), analyze
properties of the quiet Sun and their height variation or to learn
about the physics of (magneto-)convection from direct comparison with
numerical simulations.

This work may be improved upon by using new observations with even
higher spatial resolution or more spectral features. Existing
inversion codes such as NICOLE have the capability of incorporating
lines from different elements and even to combine LTE and NLTE
lines. It should be possible, in principle, to combine photospheric
and chromospheric lines with the aim of producing a 3D model like this
but spanning a much greater range of heights, all the way up to the
middle chromosphere. The main limitation, however, is that we
currently lack adequate instrumentation to produce such
high-resolution multi-wavelength observations. Future planned
ground-based telescopes such as the ATST \citep{KRH+02} or the EST
\citep{EST10} or space-born observatories such as the plan~B option of
Solar-C (Hinode's successor, still in early planning stages) would
provide an enormous leap in our ability to acquire the necessary data.

%

\begin{acknowledgements}

The author thanks Nataliya Shchukina and Javier Trujillo Bueno for
kindly providing the departure coefficients used for the NLTE
correction in this work.

Hinode is a Japanese mission developed and launched by ISAS/JAXA,
collaborating with NAOJ as a domestic partner, NASA and STFC (UK) as
international partners. Scientific operation of the Hinode mission is
conducted by the Hinode science team organized at ISAS/JAXA. This team
mainly consists of scientists from institutes in the partner
countries. Support for the post-launch operation is provided by JAXA
and NAOJ (Japan), STFC (U.K.), NASA, ESA, and NSC (Norway). 

The author thankfully acknowledges the technical expertise and
assistance provided by the Spanish Supercomputing Network (Red
Española de Supercomputación), as well as the computer resources used:
the LaPalma Supercomputer, located at the Instituto de Astrofísica de
Canarias.

Financial support by the Spanish Ministry of Science and Innovation
through project AYA2010-18029 (Solar Magnetism and Astrophysical
Spectropolarimetry) is gratefully acknowledged.
\end{acknowledgements}

\bibliographystyle{aa}

\bibliography{../bib/aamnem99,../bib/articulos}


\end{document}